\documentstyle[twoside,epsfig]{hsproc2}
\def\runauthor{G. Ecker}
\def\shorttitle{Anomalous magnetic moment}
\newcommand{\beq}{\begin{equation}}
\newcommand{\eeq}{\end{equation}}
\newcommand{\beqa}{\begin{eqnarray}}
\newcommand{\eeqa}{\end{eqnarray}}
\newcommand{\beqan}{\begin{eqnarray*}}
\newcommand{\eeqan}{\end{eqnarray*}}
\newcommand{\ba}{\begin{array}}
\newcommand{\ea}{\end{array}}

\newcommand{\ol}{\overline}

\newcommand{\nn}{\nonumber \\}
\newcommand{\bea}{\begin{eqnarray}}
\newcommand{\eea}{\end{eqnarray}}

\newcommand{\hepph}[1]{{\tt hep-ph/#1}}

\newcommand{\mrm}{\mathrm}

\begin{document}
\rightline{UWThPh-2002-34}
\vspace*{.3cm} 
\title{Two-pion contribution \\to the muon magnetic moment
\footnote{Work supported in part by RTN, EC-Contract No.
HPRN-CT-2002-00311 (EURIDICE).}}
\author{G. Ecker}{Institut f\"ur Theoretische Physik, Universit\"at
 Wien\\ Boltzmanngasse 5, A-1090 Wien, Austria}
\abstract{The two-pion contribution to hadronic vacuum polarization
  can be extracted from $\tau$ decay data when isospin violating and
  radiative corrections are taken into account. When the
  dominant corrections are applied to the photon-inclusive decay
  $\tau^- \to \nu_\tau \pi^- \pi^0 [\gamma]$, one obtains a shift
  $\Delta a_\mu = (- 12.0 \pm 2.6)\times 10^{-10}$ for the anomalous
  magnetic moment of the muon. The shift appears to be too
  small to reconcile  the determinations of hadronic vacuum 
  polarization from existing $\tau$ and $e^+e^-$ data. The
  reliability of electromagnetic corrections in the photon-inclusive 
  $\tau$ decay is examined.}
\section{Status of $g_\mu - 2$}
The magnetic moment of a particle with spin $\vec S_p$,
\begin{equation} 
\vec \mu_p = g_p \displaystyle\frac{e \hbar}{2 m_p c} \vec S_p ~,
\end{equation} 
is expressed in terms of its gyromagnetic factor $g_p$. With polarized 
muons in a properly tuned storage ring \cite{YKS02}, the spin 
precession frequency is directly proportional to the anomalous 
magnetic moment $a_\mu=(g_\mu - 2)/2$. On the basis of the most 
recent experimental result from Brookhaven \cite{BNL02}, the present 
world average is
\begin{equation} 
a_\mu^{\rm exp} = (11659203 \pm 8) \times 10^{-10}~.
\label{eq:exp}
\end{equation} 

The standard model prediction for $a_\mu$ consists of three parts:
\begin{equation} 
a_\mu^{\rm SM} = a_\mu^{\rm QED} + a_\mu^{\rm weak} +
                     a_\mu^{\rm had}~. 
\end{equation} 
The first two contributions are known with very high accuracy 
\cite{EdR02}: 
\begin{eqnarray} 
a_\mu^{\rm QED} &=& (11658470.6 \pm 0.3) \times 10^{-10}\nn
a_\mu^{\rm weak} &=& (15.2 \pm 0.1) \times 10^{-10}~.
\end{eqnarray} 
To set the stage, we may compare the experimental result with the
combined electroweak contribution,
\begin{equation} 
a_\mu^{\rm exp} - a_\mu^{\rm QED+weak} = (717 \pm 8)
\times 10^{-10}~,
\end{equation} 
a difference of 90 standard deviations. 

Although the primary
motivation of the Brookhaven experiment was to test the weak
contribution $a_\mu^{\rm weak}$, the main issue at present for a
meaningful comparison between theory and experiment is to
understand the hadronic contribution $a_\mu^{\rm had}$:
\begin{equation} 
a_\mu^{\rm had} = a_\mu^{\rm had,LO} + 
\underbrace{a_\mu^{\rm had,HO}}_{-10.0\pm 0.6} +
\underbrace{a_\mu^{\rm had,LBL}}_{8\pm 4}~. 
\label{eq:amuhad}
\end{equation}
\begin{figure} 
\centerline{\epsfig{file=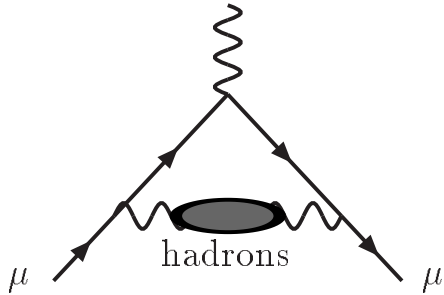,width=6cm}}
\caption{Contribution of lowest-order hadronic vacuum polarization to 
$a_\mu$.}
\label{fig:hvp}
\end{figure} 
By far the most important hadronic contribution is due to vacuum
polarization at lowest order in $\alpha$ (shown in 
Fig.~\ref{fig:hvp}):
\begin{equation} 
a_\mu^{\rm had,LO} = a_\mu^{\rm hvp}~.
\end{equation} 
The most recent value for the higher-order hadronic 
contribution \cite{Krause} is displayed in
Eq.~(\ref{eq:amuhad}) (all values for $a_\mu$ are given in units of 
$10^{-10}$ from now on). The hadronic light-by-light contribution 
$a_\mu^{\rm had,LBL}$ still carries a large theoretical uncertainty
\label{EdR02} but at least the sign is now established \cite{KN02,
KNPR02}.

Most of the hadronic vacuum polarization contribution to $a_\mu$
originates at rather low energies (about 70 \% 
for $t \le 0.8$ GeV$^2$). Therefore, nonperturbative methods are
needed together with experimental input. The most recent evaluation,
using either only $e^+e^-$ data or including $\tau$ decay data 
(for the two- and four-pion channels), finds \cite{DEHZ02}
\begin{equation} 
a_\mu^{\rm hvp} = \left\{\begin{array}{ll}
684.7 \pm 7.0 & \quad [e^+ e^-] \\
701.9 \pm 6.2 & \quad [~\tau~] 
\end{array} \right.~.
\label{eq:eetau}
\end{equation}

Comparison of the total standard model contribution with the
experimental result (\ref{eq:exp}) leads to \cite{DEHZ02}
\begin{equation} 
a_\mu^{\rm exp} - a_\mu^{\rm SM} =\left\{\begin{array}{ll}
33.9 \pm 11.2 & (3.0 ~\sigma) ~~[e^+ e^-]\\
16.7 \pm 10.7 & (1.6 ~\sigma) ~~[~\tau~]
\end{array} \right. ~.
\label{eq:comp}
\end{equation} 
The two determinations of $a_\mu^{\rm hvp}$ in Eq.~(\ref{eq:eetau})
differ by more than four standard deviations \cite{DEHZ02}. This 
discrepancy could be of experimental origin or the theoretical 
analysis might be inaccurate or incomplete. It is the main purpose 
of this talk to discuss whether the discrepancy could be due to an 
underestimate of isospin violation.

\section{Hadronic vacuum polarization and isospin violation}
The contribution of hadronic vacuum polarization at $O(\alpha^2)$
to $a_\mu$ (Fig.~\ref{fig:hvp}) is given by \cite{GdR69}
\begin{equation} 
a_\mu^{\rm hvp}=\displaystyle\int_{4
M_\pi^2}^{\infty} dt K(t) \sigma_0(e^+ e^- \to  ~\mrm{hadrons})(t)
\end{equation} 
with a smooth kernel $K(t)$ concentrated at low energies. I
discuss here only the two-pion contribution that accounts for
73 \% of $a_\mu^{\rm hvp}$.

The two-pion contribution can also be extracted from $\tau$ decay. 
In the isospin limit,
\begin{equation} 
\sigma_0(e^+ e^- \to \pi^+ \pi^-)(t)= h(t) 
\displaystyle\frac{d \Gamma(\tau^-\to \pi^0 \pi^- \nu_\tau)}{dt}
\label{eq:CVC}
\end{equation} 
with a known kinematic function $h(t)$. At the level of accuracy
needed to match the present experimental precision, a systematic
account of isospin violation including electromagnetic corrections is
required. I report here on a recent analysis of those corrections
\cite{CEN01,CEN02}.

We expect the size of isospin violating corrections to lie somewhere 
between
$$
\frac{M_{\pi^+}^2-M_{\pi^0}^2}{M_{\rho}^2}=2\times 10^{-3} 
\quad \mrm{and} \quad
\frac{M_{\pi^+}^2-M_{\pi^0}^2}{\ol{M_{\pi}^2}}=0.067~.
$$
In a first step one integrates out all heavy fields with masses 
$> m_\tau$. This generates an electroweak short-distance correction 
factor for semihadronic $\tau$ decays \cite{MS88} $S_{EW}=1.0194$ 
(in the $\ol{MS}$ scheme). The second and final step is then to 
calculate the isospin violating corrections in the theory with light 
fields only.

The CVC relation (\ref{eq:CVC}) gets modified in the presence of isospin
violation:
\begin{equation} 
\sigma_0(t) =  h(t)\displaystyle\frac{d \Gamma(\tau^-\to \pi^0 \pi^-
\nu_\tau)}{dt} \, \displaystyle\frac{R_{IB} (t)}{S_{EW}}~, 
\end{equation}
with an isospin breaking correction function
\begin{equation} 
 R_{IB} (t) =    
\displaystyle\frac{1}{G_{EM}(t)} \displaystyle\frac{\beta^3_{\pi^+ 
\pi^-}(t)}{\beta^3_{\pi^0\pi^-}(t)} \left|\displaystyle\frac{F_V(t)}
{f_+(t)}\right|^2 ~.
\end{equation}
The phase space correction factor \cite{CK01}
\begin{equation} 
\frac{\beta^3_{\pi^0 \pi^-}(t)}{\beta^3_{\pi^+\pi^-}(t)}=
1 + \frac{3(M_{\pi^+}^2-M_{\pi^0}^2)}{t-4 M_\pi^2}+ 
O[(M_{\pi^+}^2-M_{\pi^0}^2)^2]
\eeq
is especially important near threshold. The ratio of form factors 
($F_V(t)$ in  
$e^+e^-$ annihilation, $f_+(t)$ in $\tau$ decay) is mainly sensitive 
to $\rho-\omega$ mixing and (to a lesser extent) to the width difference
$\Gamma_{\rho^+}-\Gamma_{\rho^0}$ \cite{DEHZ02,CEN02}.

The main task is the calculation of radiative corrections that go into
the function $G_{\rm EM}(t)$. As usual, the radiative
corrections consist of two parts and only the sum is infrared finite
and well defined: the exclusive rate with one-loop corrections
\cite{CEN01} and the radiative rate. I concentrate here on the
calculation of the radiative rate \cite{CEN02}, i.e., 
$\Gamma(\tau^-\to \pi^0 \pi^- \nu_\tau \gamma)$ under ALEPH conditions
\cite{ALEPH}, with photons of all energies included. 

To describe this decay, we have used a gauge invariant chiral
resonance model with the following features \cite{CEN02}:
\begin{itemize}
\item Low's theorem (leading and subleading terms) is manifestly
  satisfied in terms of an explicit representation for the pion form
  factor $f_+(t)$ \cite{GP97}.
\item The amplitude exhibits the correct low-energy behaviour to
  $O(p^4)$. 
\item The low-energy amplitude is extended into the resonance region
  using the standard chiral resonance Lagrangian \cite{EGPR89}. The 
  implicit assumption is that $\rho$ and (to a lesser extent) $a_1$
  exchange, which contribute already at $O(p^4)$, are the dominant 
  mechanisms at all accessible energies.
\end{itemize} 

For photon energies $E_\gamma < 100$ MeV (in the $\tau$ rest frame), 
the rate is dominated by bremsstrahlung (leading Low approximation). 
However, under ALEPH conditions with all photons included, the
bremsstrahlung approximation is not sufficient. The infrared finite 
sum of loop-corrected and radiative rate translates into the function
$G_{\rm EM}(t)$ shown in Fig.~\ref{fig:GEM}.
\begin{figure} 
\centerline{\epsfig{file=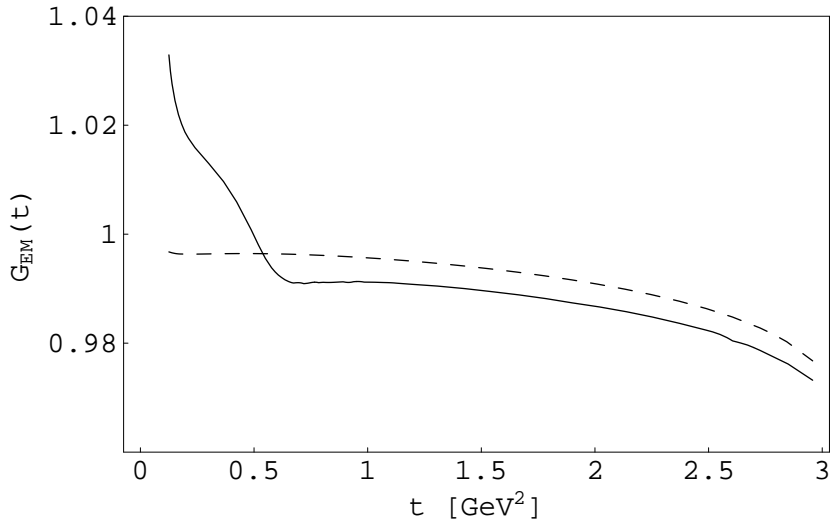,height=7cm}}
\caption{Electromagnetic correction function $G_{\rm EM}(t)$: the 
full curve corresponds to the complete radiative amplitude, the dashed 
curve is based on the leading Low approximation (bremsstrahlung) for the
radiative mode.}
\label{fig:GEM}
\end{figure}

We are now ready to calculate the total shift in $a_\mu^{\rm hvp}$ due
to isospin violation in the two-pion channel:
\begin{equation} 
\Delta a_\mu^{\rm hvp}=\displaystyle\int_{4 M_\pi^2}^{t_{\rm max}} 
dt K(t) \left[h(t) \frac{ d \Gamma_{\pi \pi [\gamma]}}{d t} \right] 
\times \left(\frac{R_{\rm IB} (t) }{S_{\rm EW}} \, - \, 1 \right)~.
\end{equation} 
In Table \ref{tab:shift} the various contributions and the total
shifts are displayed for two values of $t_{\rm max}$: electroweak 
short-distance correction ($S_{\rm EW}$), threshold correction (KIN), 
radiative corrections (EM) and form factor ratio (FF). Obviously, the 
low-energy region below 1 GeV$^2$ is dominating.
\begin{table}
\begin{center}
\begin{tabular}{|c|c|c|c|c|c|}
\hline 
 & & &  & & \\ 
 $t_{\rm max}$(GeV$^2$) &  $S_{\rm EW}$  &  KIN    
& EM  &  FF & total   \\ 
 & & &  & & \\
\hline
 & & &  & & \\
1 & - 9.5   &   - 7.5  & - 1.1 & 6.1 $\pm$ 2.6 
  & - 11.9 $\pm$ 2.6   \\
 & & &  & & \\
3 & - 9.7  &   - 7.5  & - 1.0 &
 6.1 $\pm$ 2.6  &  - 12.0 $\pm$ 2.6 \\
 & & &  & & \\
\hline 
\end{tabular}
\caption{Contributions to $\Delta a_\mu^{\rm hvp}$  from 
various sources of isospin violation (in units of $10^{-10}$) for 
two different values of 
$t_{\max}$ (in units of GeV$^2$; $t_{\max}\le m_\tau^2$).} 
\label{tab:shift}
\end{center}
\end{table}

I have only listed our error estimate for the form factor
ratio because the uncertainty associated with some electromagnetic 
low-energy
constants appearing in $G_{\rm EM}(t)$ is much smaller \cite{CEN02}.
The short-distance and kinematic corrections are model independent 
(although there could be higher-order corrections in $S_{\rm EW}$). 
The form factor ratio is dominated by  $\rho-\omega$ mixing taken 
directly from the most recent experimental analysis \cite{CMD02}.

The total isospin violating correction goes in the right direction
towards reconciling the $\tau$ with the $e^+e^-$ data but the
shift seems to be too small in absolute magnitude. While the signs and
magnitudes of the shifts caused by the short-distance correction, the
kinematic threshold effect and the form factor ratio are well 
understood, the small electromagnetic shift deserves further discussion.

A first observation is that loop and radiative contributions tend
to interfere destructively (for reasonable infrared cutoffs because
only the sum is infrared finite). A more instructive exercise is the
comparison with the radiative corrections for the inclusive rate
$\Gamma(\tau^-\to d\, \ol{u}\, \nu_\tau [\gamma])$ at the quark level,
calculated some time ago by Braaten and Li \cite{BL90}. Translating 
their result into a shift in $a_\mu$, one finds
\begin{equation} 
\Delta a_\mu^{\rm EM,quark}=1.6~,
\end{equation} 
of opposite sign but similar magnitude as in Table \ref{tab:shift}.
There is no fundamental reason why the inclusive result at the quark 
level should agree with the exclusive two-pion result. Nevertheless, 
performing the radiative corrections for the two-pion mode in the 
leading Low approximation for the radiative mode (independent of any 
resonance contributions to the decay amplitude), one obtains
accidentally exactly the same value:
\begin{equation} 
\Delta a_\mu^{\rm EM,Low}=1.6~.
\end{equation} 
Note that the shift in the inclusive case and in the bremsstrahlung 
approximation for the exclusive channel goes in the ``wrong''
direction increasing the discrepancy with the $e^+e^-$ result.

What is then the origin of the ``correct'' sign of the shift 
$\Delta a_\mu^{\rm EM}$ in Table \ref{tab:shift}? It turns out that
the difference to the bremsstrahlung value is completely due to the
subleading terms in the Low expansion for the radiative amplitude 
proportional to the derivative $d f_+(t)/dt$ of the pion form
factor \cite{CEN02}. In other words, the full curve in
Fig.~\ref{fig:GEM} could not be distinguished from the curve based
only on the first two terms in the Low expansion. 
To the extent that the pion form factor is known
experimentally, the shift $\Delta a_\mu^{\rm EM}$ is therefore model
independent and certainly independent of details of the
resonance exchange model for the radiative decay
$\tau^-\to \pi^0 \pi^- \nu_\tau \gamma$. The result depends only on
the (shape of the) pion form factor.

\section{Conclusions}
The total shift of $a_\mu^{\rm hvp}$ due to isospin violation and
radiative corrections in the two-pion channel,
\begin{equation} 
\Delta a_\mu^{\rm hvp} = - 12.0 \pm 2.6~,
\end{equation}
agrees well with a similar more data-oriented analysis of Davier et
al. \cite{DEHZ02}. Performing the shift for $a_\mu^{\rm hvp}$
extracted from the two-pion decay of the $\tau$, one
arrives at the results \cite{DEHZ02} already displayed in 
Eq.~(\ref{eq:comp}):
\begin{equation}
a_\mu^{\rm exp} - a_\mu^{\rm SM} =\left\{\begin{array}{ll}
33.9 \pm 11.2 & (3.0 ~\sigma) ~~[e^+ e^-]\\
16.7 \pm 10.7 & (1.6 ~\sigma) ~~[~\tau~]
\end{array} \right. ~.
\end{equation}  

Before drawing any far-reaching conclusions about possible evidence for 
new physics, the reason for the discrepancy between the two 
determinations of $a_\mu^{\rm hvp}$ must be understood. One possible 
uncertainty has to do with the experimental procedure of applying 
radiative corrections to the $e^+e^-$ data where the corrections are 
much bigger than in the $\tau$ decay. If both the $e^+e^-$ result 
and the raw $\tau$ data
were correct isospin violation would have to be more than twice as
big as calculated \cite{CEN02}. In view of the discussion presented
here, especially on the natural size of electromagnetic corrections 
in the $\tau$ decay to two pions, I consider such a drastic 
underestimate very unlikely.

For the time being and pending clarification of the discrepancy
between $e^+e^-$ and $\tau$-based extractions of $a_\mu^{\rm hvp}$,
the standard model prediction for the anomalous magnetic moment of the
muon is in good shape.

\vspace*{.5cm} 

\section*{Acknowledgments}
I wish to congratulate G. Martinsk\'a, J. Urb\'an, S. Vok\'al and
their team for the excellent organization of 
Hadron Structure '02. I also thank V. Cirigliano and H. Neufeld for 
a very pleasant collaboration.


\vspace*{.5cm} 

\begin{thebibliography}{99}
\bibitem{YKS02}
Y.K. Semertzidis: \hepph{0211038} and references therein.
\bibitem{BNL02}
\refer{G.W. Bennett et al. (Muon $g-2$ Coll.)}{Phys. Rev. Lett.}{89}
{2002}{101804.}
\bibitem{EdR02}
E. de Rafael: \hepph{0208251}.
\bibitem{Krause}
\refer{B. Krause}{Phys. Lett.}{B 390}{1997}{392.}
\bibitem{KN02}
\refer{M. Knecht and A. Nyffeler}{Phys. Rev.}{D 65}{2002}{0730334.}
\bibitem{KNPR02}
\refer{M. Knecht, A. Nyffeler, M. Perrottet and E. de Rafael}
{Phys. Rev. Lett.}{88}{2002}{071802.}
\bibitem{DEHZ02}
M. Davier, S. Eidelman, A. H\"ocker and Z. Zhang: \hepph{0208177}.
\bibitem{GdR69}
\refer{M. Gourdin and E. de Rafael}{Nucl. Phys.}{B 10}{1969}{667.}
\bibitem{CEN01}
\refer{V. Cirigliano, G. Ecker and H. Neufeld}{Phys. Lett.}
{B 513}{2001}{361.}
\bibitem{CEN02}
\refer{V. Cirigliano, G. Ecker and H. Neufeld}{JHEP}{08}{2002}{002.}
\bibitem{MS88}
\refer{W.J. Marciano and A. Sirlin}{Phys. Rev. Lett.}{61}{1988}{1815;} 
ibid. {\bf 71} (1993) 3629.
\bibitem{CK01}
\refer{H. Czyz and J.H.  K\"uhn}{Eur. Phys. J.}{ C18}{2001}{497.}
\bibitem{ALEPH}
\refer{R. Barate et al. (ALEPH)}{Z. Physik}{C 76}{1997}{15.}
\bibitem{GP97}
\refer{F. Guerrero and A. Pich}{Phys. Lett.}{B 412}{1997}{382.}
\bibitem{EGPR89}
\refer{G. Ecker, J. Gasser, A. Pich and E. de Rafael}{Nucl. Phys.}
{B 321}{1989}{311.}
\bibitem{CMD02}
\refer{R.R. Akhmetshin et al. (CMD-2)}{Phys. Lett.}{B 527}{2002}{161.}
\bibitem{BL90}
\refer{E. Braaten and C.S. Li}{Phys. Rev.}{D 42}{1990}{3888.}

\end{thebibliography}
\end{document}